\documentclass[aps,pre,floatfix,12pt,showkeys]{revtex4-1}
\usepackage{amsfonts}
\usepackage[T2A]{fontenc}
\usepackage[utf8]{inputenc}
\usepackage{amsmath}
\usepackage{amssymb}
\usepackage[english]{babel}
\usepackage{graphicx}
\usepackage{mathtools}

\begin{document}

\title{Analytical solution of linearized equations of the Morris-Lecar neuron model at large constant stimulation}

\author{A.V. Paraskevov$^{1,2}$, T.S. Zemskova$^{3,4}$}

\affiliation{$^1$Institute for Information Transmission Problems, 127051 Moscow, Russia\\$^2$National Research Centre "Kurchatov Institute", 123182 Moscow, Russia\\$^3$Ecole Polytechnique, 91128 Palaiseau, France\\$^4$Moscow Institute of Physics and Technology (National Research University), 141700 Dolgoprudny, Russia}

\begin{abstract}
\begin{center}
\textbf{Abstract}
\end{center}
\noindent The classical biophysical Morris-Lecar model of neuronal excitability predicts that upon stimulation of the neuron with a sufficiently large constant depolarizing current there exists a finite interval of the current values where periodic spike generation occurs. Above the upper boundary of this interval, there is four-stage damping of the spike amplitude: 1) minor primary damping, which reflects a typical transient to stationary dynamic state, 2) plateau of nearly undamped periodic oscillations, 3) strong damping, and 4) reaching a constant asymptotic value of the neuron potential. We have shown that in the vicinity of the asymptote the Morris-Lecar equations can be reduced to the standard equation for exponentially damped harmonic oscillations. Importantly, all coefficients of this equation can be explicitly expressed through parameters of the original Morris-Lecar model, enabling direct comparison of the numerical and analytical solutions for the neuron potential dynamics at later stages of the spike amplitude damping.

\bigskip

\end{abstract}

\maketitle

\textbf{1. Introduction}

\bigskip

The Morris-Lecar (ML) model \cite{ML1981,MIT1998} is a classical biophysical model of spike generation by the neuron, which takes into account the dynamics of voltage-dependent ion channels and realistically describes the spike waveform. The model predicts that upon stimulation of the neuron with sufficiently large constant depolarizing current $I_{stim}$, there exists a finite interval of $I_{stim}$ values where periodic spike generation occurs \cite{MIT1998,Rob2004,Boris2005,Tsu2006,BC2012,Liu2014}. Numerical simulations have shown that in the ML model the cessation of periodic generation of spikes above the upper boundary of this interval (i.e. at $I_{stim}$ > $I_{max}$ in \mbox{Fig. 1}) occurs through a damping of the spike amplitude, arising with a delay inversely proportional to the value of $I_{stim}$ \cite{BConf2018}. In particular, the damped dynamics can be divided into four successive stages: 1) minor primary damping, which reflects a typical transient to stationary dynamic state, 2) plateau of nearly undamped periodic oscillations, which determines the aforementioned delay, 3) strong damping, and 4) reaching constant stationary asymptotic value $V_{st}$ of the neuron potential. This dynamic behavior of the ML model is qualitatively the same for the 1st and 2nd types of neuronal excitability.

In this paper, we have found a way to linearizing the ML model equations in the vicinity of the asymptote $V_{st}$. The resulting equations have been then reduced to an inhomogeneous Volterra integral equation of the second kind. In turn, the latter has been transformed into an ordinary differential equation of the second order with a time-dependent coefficient at the first-order derivative. As this time dependence was just an exponential decay with the small pre-exponential factor, we considered its asymptotic value and analytically solved the final equation. In order to verify the analytical solution found, we have compared it with the numerical solution obtained using the standard MATLAB tools for systems of ordinary differential equations (see the Supplementary Material, which contains the MATLAB scripts and generated data used for the Figures).

As the result, we have accurately shown that the linearized system of equations of the ML model can be reduced to the standard equation of exponentially damped harmonic oscillations for the neuron potential. Since all coefficients of this equation are explicitly expressed through parameters of the original ML model, one can directly (i.e. without any fitting) compare the numerical and analytical solutions for dynamics of the neuron potential at last two stages of the spike amplitude damping (left graphs in \mbox{Fig. 2} and \mbox{Fig. 3}). The results allow a quantitative study of the applicability boundary of ordinary bifurcation analysis that implies exponential dynamics.

Finally, it should be noted that a similar effect of delayed damped oscillations of the neuronal potential has been previously reported for the ML model with the 2nd excitability type and the stimulating current value, which is just below the lower boundary of sustained spiking \cite{PRE2007,JMB2013}. Emphasize that these findings are only applicable, first, for small stimulating current values lying before the region of sustained oscillations (below $I_{min}$ in \mbox{Fig. 1}). In turn, we consider the case of a large stimulating current above that region, i.e., at a different stationary point of the ML model. Second, the results \cite{PRE2007,JMB2013} are only valid for the excitability type 2 defined by an abrupt occurrence of high-frequency sustained oscillations at finite minimal value $I_{min}$ of constant stimulating current (see Fig. S1 in the Supplementary Material). For the type 1, where sustained oscillation frequency starts as a continuous function of the stimulating current (\mbox{Fig. 1}), the damped oscillations below $I_{min}$ do not arise in principle, and the results \cite{PRE2007,JMB2013} are not applicable. On the contrary, in the case of large stimulating current considered in this paper, the model behavior is universal for both the 1st and 2nd types of excitability so that the results are also universal.

\begin{figure}[!t]
\centering
\includegraphics[width=0.8\textwidth]{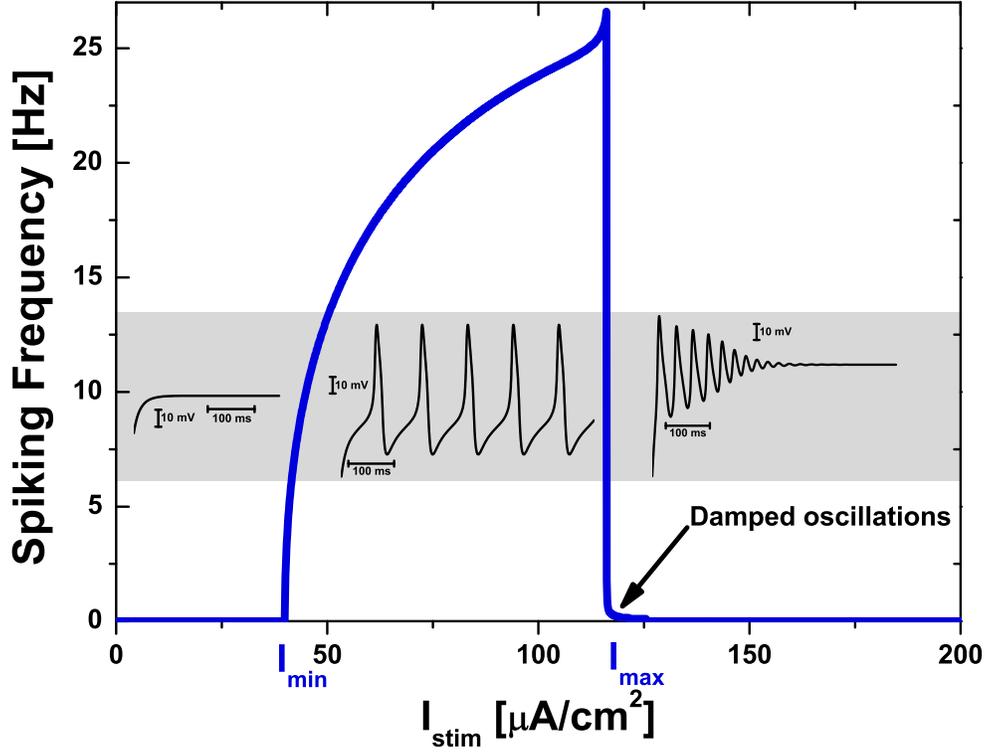}
\caption{Dependence of spike generation frequency (determined as the number of spikes divided by the time interval of 20000 ms) on constant stimulating current $I_{stim}$ and, on the gray inset, typical examples of dynamics of the neuron potential in the corresponding ranges of $I_{stim}$ values for the Morris-Lecar model with the 1st excitability type \cite{MIT1998}. Spikes are characteristic pulses of the neuron potential (see the gray inset in the range from $I_{min}=40$ $\mu$A/cm$^{2}$ to $I_{max}=116.1$ $\mu$A/cm$^{2}$).}
\label{Fig1}
\end{figure}

\bigskip

\textbf{2. Standard Morris-Lecar model}

\bigskip

As phase plane analysis of the ML model is extensively described in textbooks (e.g., \cite{Koch99,CCB2002,Izh07,MathNeuro2010,Prin2011,ND2014}), we omit it and provide only basic facts on the model, and its formal description.

Qualitatively, the classical two-dimensional ML model \cite{ML1981,MIT1998} (cf. \cite{Gonz2014}) couples dynamics of the transmembrane potential $V$ of the neuron with dynamics of the transmembrane conductance $w$ of potassium ions. Spikes represent characteristic pulses of $V$ (see the gray inset in \mbox{Fig. 1}, in the range from $I_{min}$ to $I_{max}$). In the ML model the rate of change of $V$ depends on the current value of $w$ in such a way that dynamics of $w$ provides a negative feedback with respect to the dynamics of $V$. In turn, the rate of change of $w$ is proportional to the difference between the current value of $w$ and some "asymptotic" value $w_{\infty}$, which nonlinearly depends on $V$. As a result, $w$ tends to reach $w_{\infty}$, which is changed itself in time due to dynamics of $V$. If one neglects the relaxation of $w$ to $w_{\infty}$, i.e., assumes that this occurs instantly, then the generation of spikes in the ML model does not happen. The upper value of the stimulating current, $I_{max}$, above which the continuous periodic generation of spikes stops, corresponds to the onset of a relatively fast relaxation of $w$ to $w_{\infty}$.

Quantitatively, the standard ML model equations for dynamics of the neuronal potential $V$ and for relaxation dynamics of the normalized conductance $w$ of potassium ions are given by
\begin{equation}
\begin{dcases}
C_{m}dV/dt = -I_{ion}(V,w)+I_{stim},\\
dw/dt = (w_{\infty}(V)-w)/\tau(V),
\end{dcases}\label{ML1}
\end{equation}
where the total sum of ion currents
\begin{equation}
I_{ion}(V,w) = g_{Ca}m_{\infty}(V)(V-V_{Ca})+g_{K}w(V-V_{K})+g_{L}(V-V_{L}), \label{ML2}
\end{equation}
$I_{stim}$ is an external stimulating current, and the constituent functions
\begin{align}
m_{\infty}(V) &  =\frac{1}{2}\left[  1+\tanh((V-V_{1})/V_{2})\right], \label{ML3} \\
w_{\infty}(V) &  =\frac{1}{2}\left[  1+\tanh((V-V_{3})/V_{4})\right], \label{ML4}\\
\tau(V) &  =\tau_{\max}/\cosh((V-V_{3})/(2V_{4})). \label{ML5}
\end{align}

For numerical simulations shown in Figures 1-3 we have used the following values of the ML model parameters corresponding the 1st neuronal excitability type \cite{MIT1998}: $C_{m}=20$ $\mu$F/cm$^{2}$, $g_{Ca}=4$ mS/cm$^{2}$, $g_{K}=8$ mS/cm$^{2}$, $g_{L}=2$ mS/cm$^{2}$, $V_{Ca}=120$ mV, $V_{K}=-84$ mV, $V_{L}=-60$ mV, $V_{1}=-1.2$ mV, $V_{2}=18$ mV, $V_{3}=12$ mV, $V_{4}=17.4$ mV, $\tau_{\max}=14.925$ ms. These parameters result in the resting potential value $V_{rest}=-59.47$ mV, which is the solution of equation $I_{ion}(V,w_{\infty}(V))=0$ and is very close to $V_{L}$ value. Supplementary Figures S1-S2 show results for the ML model of the 2nd excitability type \cite{MIT1998}, for which $g_{Ca}=4.4$ mS/cm$^{2}$, $V_{3}=2$ mV, $V_{4}=30$ mV, $\tau_{\max}=25$ ms, and all the rest parameters are the same as those for the 1st type. In turn, these parameters result in $V_{rest}=-60.85$ mV.

The initial conditions for all numerical simulations of the ML model in this paper were as follows: $V(t=0)=V_{rest}$, $w(t=0)=w_{\infty}(V_{rest})$.

\bigskip

\textbf{3. Linearization of the Morris-Lecar equations at large constant stimulation}

\bigskip

In what follows, we consider the case $I_{stim}>I_{max}$ and seek a solution for the potential in the form $V(t)=V_{st}+U(t)$, where $V_{st}$ is the stationary potential value determined from equation $I_{ion}(V_{st},w_{\infty}(V_{st}))=I_{stim}$ and $U(t\rightarrow+\infty)=0$. In addition, we assume that for any moment of time $t$ the condition $\left\vert U(t)\right\vert \ll\left\vert V_{st}\right\vert$ holds. Given this, we expand $w_{\infty}(V)$ and $m_{\infty}(V)$ into a Taylor series up to the linear term with respect to $U$:
\begin{align*}
w_{\infty}(V_{st}+U)  &  \approx w_{\infty}(V_{st})+\frac{dw_{\infty}(V_{st})}{dV}U=a+bU,\\
m_{\infty}(V_{st}+U)  &  \approx m_{\infty}(V_{st})+\frac{dm_{\infty}(V_{st})}{dV}U=p+qU,
\end{align*}
where $a=w_{\infty}(V_{st})$, $b=dw_{\infty}(V_{st})/dV$, $p=m_{\infty}(V_{st})$, $q=dm_{\infty}(V_{st})/dV$.

Next, we assume that $\tau(V)\approx\tau(V_{st})\approx\tau_{\max}\equiv\tau$. The assumption is quite important for the linearizing and is based on preliminary numerical simulations showing that the value of $I_{max}$ does not change substantially (at maximum, for a few percents) with this assumption, regardless to the neuronal excitability type (see Fig. S1).

After that, we exactly solve the linearized ML equation on $w$,
\begin{equation}
\begin{dcases}
\tau dw/dt = a+bU(t)-w,\\
w(t=t_{0})=w_{0}.
\end{dcases}
\end{equation}
Its general solution has form
\begin{equation}
w(t)=a+W_{0}(t)+b\exp(-\frac{t}{\tau}){\displaystyle\int\limits_{t_{0}}^{t}}\exp(\frac{t^{\prime}}{\tau})\frac{U(t^{\prime})}{\tau}dt^{\prime},\label{wt}
\end{equation}
where $W_{0}(t)=(w_{0}-a)\exp(-(t-t_{0})/\tau)$.

We find value $w(t_{0})=w_{0}$ in a local extremum point of the potential $V(t=t_{0})=V_{0}=V_{st}+U_{0}$, which is determined by condition $\frac{dV}{dt}(t=t_{0})=0$. One gets
\begin{equation}
w_{0}=\frac{I_{stim}-g_{Ca}m_{\infty}(V_{0})(V_{0}-V_{Ca})-g_{L}(V_{0}-V_{L})}{g_{K}(V_{0}-V_{K})}.\label{w0}
\end{equation}
Further, writing explicitly the equation on $U$ and neglecting nonlinear terms, we obtain a linear integro-differential equation for the potential $U$,
\begin{equation}
\frac{dU}{dt}=-G(t)-A(t)U-B\exp(-\frac{t}{\tau}){\displaystyle\int\limits_{t_{0}}^{t}}\exp(\frac{t^{\prime}}{\tau})\frac{U(t^{\prime})}{\tau}dt^{\prime},
\end{equation}
where coefficients $A(t)$, $B$, and $G(t)$ are as follows (cf. \cite{JMB2013}):
\begin{align*}
A(t)  &  =[g_{Ca}(p+q(V_{st}-V_{Ca}))+g_{K}(a+W_{0}(t))+g_{L}]/C_{m}\equiv A+A_{0}(t),\\
A  &  =[g_{Ca}(p+q(V_{st}-V_{Ca}))+g_{K}a+g_{L}]/C_{m},\\
A_{0}(t)  &  =g_{K}W_{0}(t)/C_{m}\equiv A_{K}\exp(-(t-t_{0})/\tau),\text{ \ \ }A_{K}=g_{K}(w_{0}-a)/C_{m},\\
B  &  =g_{K}b\left(V_{st}-V_{K}\right)/C_{m},\\
G(t) &  =A_{0}(t)(V_{st}-V_{K})=B((w_{0}-a)/b)\exp(-(t-t_{0})/\tau).
\end{align*}
Integrating by parts, we obtain
\begin{equation}
\frac{dU}{dt}=G_{1}(t)-A_{1}(t)U(t)+B\exp(-\frac{t}{\tau}){\displaystyle\int\limits_{t_{0}}^{t}}\exp(\frac{t^{\prime}}{\tau})\frac{dU}{dt^{\prime}}dt^{\prime},
\end{equation}
where $G_{1}(t)=-G(t)+BU(t_{0})\exp(-(t-t_{0})/\tau)$ and $A_{1}(t)=A(t)+B$.

Further, given that $U(t)=U(t_{0})+{\displaystyle\int\limits_{t_{0}}^{t}}(dU/dt^{\prime})dt^{\prime}$, one can reduce the previous equation on $U$ to an integral equation for its derivative $f(t)=dU/dt$,
\begin{equation}
f(t)=G_{2}(t)+{\displaystyle\int\limits_{t_{0}}^{t}}K(t,t^{\prime})f(t^{\prime})dt^{\prime}, \label{f}
\end{equation}
where $G_{2}(t)=G_{1}(t)-A_{1}(t)U(t_{0})=G_{3}+G_{4}\exp(-(t-t_{0})/\tau)$, $G_{3}=-(A+B)U(t_{0})$, $G_{4}=U(t_{0})(-A_{K}+B)-B(w_{0}-a)/b$, and
\begin{equation}
K(t,t^{\prime})=-A_{1}(t)+B\exp(-\frac{(t-t^{\prime})}{\tau})=-(A+B)-A_{K}\exp(-\frac{(t-t_{0})}{\tau})+B\exp(-\frac{(t-t^{\prime})}{\tau}).\label{Ktt}
\end{equation}
The resulting equation \eqref{f} for $f(t)$ is an inhomogeneous Volterra integral equation of the second kind. Twice differentiating both sides of Eq. \eqref{f} with respect to $t$, we obtain that the integral equation \eqref{f} is equivalent to ordinary differential equation of the second order
\begin{equation}
\frac{d^{2}f}{dt^{2}}+\left(2\gamma+A_{K}\exp(-(t-t_{0})/\tau)\right)\frac{df}{dt}+\left(-\frac{A_{K}}{\tau}\exp(-(t-t_{0})/\tau)+\omega_{0}^{2}\right)f(t)=0,\label{feq}
\end{equation}
where constants $2\gamma=A+1/\tau$ and $\omega_{0}^{2}=(A+B)/\tau$ have been introduced.

Returning to potential $U$ and allocating the full derivative, we have
\begin{equation}
\frac{d^{2}U}{dt^{2}}+\left(2\gamma+A_{K}\exp(-(t-t_{0})/\tau)\right) \frac{dU}{dt}+\omega_{0}^{2}U=const.
\end{equation}
Assuming that potential $U(t)$ and all its derivatives tend to zero at $t\rightarrow+\infty$, one gets $const=0$. Finally, we obtain
\begin{equation}
\frac{d^{2}U}{dt^{2}}+\left(2\gamma+A_{K}\exp(-(t-t_{0})/\tau)\right)\frac{dU}{dt}+\omega_{0}^{2}U=0,\label{Ueq}
\end{equation}
with initial conditions
\begin{equation}
U(t=t_{0})=U_{0}= V_{0}-V_{st},\text{ \ \ }\frac{dU}{dt}(t=t_{0})=0.\label{IC}
\end{equation}

\begin{figure}[!t]
\centering
\includegraphics[width=\textwidth]{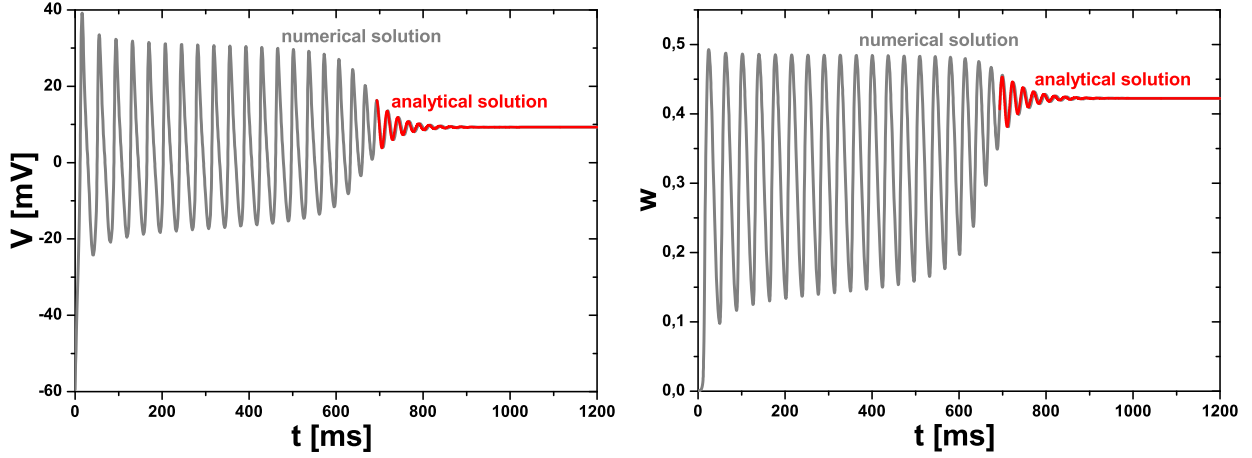}
\caption{Left graph: The gray curve is a numerical solution for dynamics of neuron potential $V(t)$ in the Morris-Lecar model with the 1st excitability type at $I_{stim}$ = 116.3 $\mu$A/cm$^{2}$ > $I_{max}$ = 116.1 $\mu$A/cm$^{2}$. The red curve is an analytical solution of the linearized system of equations of the Morris-Lecar model with initial conditions taken at the point of a local maximum of the potential ($t_{0}$ = 693.3 ms, $V_{0}$ = 16.35 mV). Right graph: The corresponding numerical (gray) and analytical (red) solutions for $w(t)$. Parameters of the analytical formulas for this example are as follows: \mbox{$V_{st}$ = 9.28 mV}, \mbox{$a = 0.42$}, \mbox{$\omega_{0}$ = 262.1 Hz}, \mbox{$\gamma$ = 21.3 Hz}, \mbox{$\omega$ = 261.2 Hz}, \mbox{$1/\tau$ = 67.2 Hz}, \mbox{$\eta = 0.08$}, \mbox{$\chi = 0.17$}, \mbox{$2\gamma/|A_{K}| = 6.78$}, \mbox{$U_{0}$ = 7.07 mV}, \mbox{$a/w_{0} = 1.04$}, \mbox{$W_{a} = 0.05$}, and \mbox{$W_{c} = -0.02$}, where the last five parameters depend on $t_{0}$ and $V_{0}$ values.}
\label{Fig2}
\end{figure}

\bigskip

\textbf{4. Analytical solution of the linearized equations}

\bigskip

Assuming that $w_{0}\approx a$ and neglecting the time-dependent parameter in Eq. \eqref{Ueq}, we arrive at
\begin{equation}
\frac{d^{2}U}{dt^{2}}+2\gamma\frac{dU}{dt}+\omega_{0}^{2}U=0.\label{Ueq2}
\end{equation}
Given the initial conditions \eqref{IC}, the solution of Eq. \eqref{Ueq2} has form
\begin{equation}
U(t)=U_{0}\exp(-\gamma(t-t_{0}))\left[\cos(\omega(t-t_{0}))+\frac{\gamma}{\omega}\sin(\omega(t-t_{0}))\right],\label{Usol}
\end{equation}
with angular frequency $\omega=\sqrt{\omega_{0}^{2}-\gamma^{2}}$ and oscillation period $T=2\pi/\omega$. This solution describes exponentially-damped harmonic oscillations and corresponds well with the numerical result (left graph in \mbox{Fig. 2} and two left graphs in \mbox{Fig. 3}, see also Fig. S2 for the excitability type 2). It is worth noting that $\omega_{0}$ and $\gamma$ are independent of $t_{0}$ and $V_{0}$. Therefore the dependencies $\omega_{0}(I_{stim})$, $\gamma(I_{stim})$, and $\omega(I_{stim})$ are relatively universal and, moreover, these can be continued in the range $I_{stim}<I_{max}$ (\mbox{Fig. 3}, right graph), though in this case there is no correspondence between the numerical and analytical solutions.

\begin{figure*}[!t]
\centering
\includegraphics[width=\textwidth]{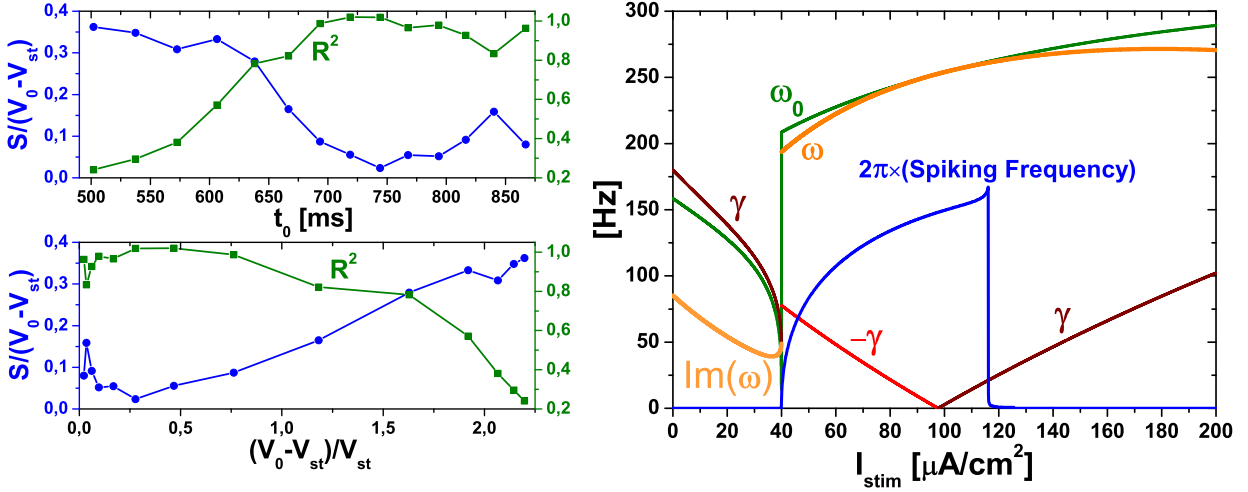}
\caption{Left graph: Quantitative evaluation of the correspondence between the numerical and analytical solutions at $I_{stim}$ = 116.3 $\mu$A/cm$^{2}$ (see left graph in \mbox{Fig. 2}): dependencies for $S=\frac{1}{n}\sum_{i=1}^{n}|V_{st}+U(t_{i})-V(t_{i})|$ (blue circles, left scale) and $R^{2}=\frac{\sum_{i=1}^{n}(V_{st}+U(t_{i})-V_{mean})^{2}}{\sum_{i=1}^{n}(V(t_{i})-V_{mean})^{2}}$ (green squares, right scale) on different values of $t_{0}$ and $V_{0}$. Here $V_{mean}=\frac{1}{n}\sum_{i=1}^{n}V(t_{i})$, and $\left\{t_{i}\right\}_{i=1}^{n}$ is the set of time moments $t_{i}>t_{0}$, for which numerical solution $V(t_{i})$ is known. As one can see from the lower graph, an approximate empirical condition of the good correspondence is $V_{0}<2V_{st}$. Right graph: Analytical dependencies of $\omega_{0}$, $\gamma$, and $\omega$ on the value of constant stimulating current $I_{stim}$, with superimposed spiking frequency from \mbox{Fig. 1}.}
\label{Fig3}
\end{figure*}

One can also obtain an explicit solution for $w(t)$ by substituting $U(t)$ into Eq. \eqref{wt}:
\begin{equation}
w(t)=a+(w_{0}-a+w_{1})\exp(-(t-t_{0})/\tau)+w_{1}\exp(-\gamma(t-t_{0}))\left[ -\cos(\omega(t-t_{0}))+H\sin(\omega(t-t_{0}))\right],\label{wsol}
\end{equation}
where $w_{1}=bU_{0}A/B$ and $H=\omega _{0}^{2}/(\omega A)-\gamma/\omega$.

Using auxiliary trigonometric transformations, one can write functions $U(t)$ and $w(t)$ in a more compact form. Denoting
\begin{equation}
\cos(\eta)=\frac{1}{\sqrt{1+(\gamma/\omega)^{2}}}=\omega/\omega _{0},\text{ \ \ }\sin(\eta)=\frac{\gamma/\omega}{\sqrt{1+(\gamma/\omega)^{2}}}=\gamma/\omega _{0},
\end{equation}
we get
\begin{equation}
U(t)=U_{0}\exp(-\gamma(t-t_{0}))\frac{\cos(\omega(t-t_{0})-\eta)}{\cos(\eta)},
\end{equation}
where $\eta=\arctan(\gamma/\omega)$ is the inverse function of $\tan(\eta)=\gamma/\omega$. This expression for $U(t)$ is completely equivalent to the previous solution \eqref{Usol}.

In turn, introducing notations
\begin{equation}
s=(1-\gamma\tau)/(\omega\tau),\text{ \ \ }\cos(\chi)=1/\sqrt{1+s^{2}},\text{ \ \ }\sin(\chi)=s/\sqrt{1+s^{2}},
\end{equation}
we obtain a compact solution for $w(t)$,
\begin{equation}
w(t)=a+W_{c}\exp(-(t-t_{0})/\tau)+W_{a}\exp(-\gamma(t-t_{0}))\sin(\omega(t-t_{0})+\chi-\eta),
\end{equation}
where $\chi=\arctan(s)$, quantity $W_{a}=(bU_{0}/(\omega\tau))\sqrt{1+A/B}$ determines the amplitude of the damped oscillations of $w(t)$ (see \mbox{Fig. 2}, right graph), and $W_{c}=(w_{0}-a)-W_{a}\sin(\chi-\eta)$.

\bigskip

\newpage

\textbf{5. Conclusion}

\bigskip

We have shown analytically, and confirmed numerically, that for the Morris-Lecar neuron model upon stimulation by large constant depolarizing current the later stages of spike amplitude damping can be accurately reduced to exponentially damped harmonic oscillations, with the frequency and damping coefficient completely determined by the original model parameters. Importantly, the obtained analytical formulas converge equally well (near the asymptote) with the numerical calculation for both the 1st and 2nd types of neuronal excitability. These formulas can be directly used to quantify deviations from the harmonic oscillations when moving away from the asymptote, i.e. with an increase in the oscillation amplitude. In other words, the results define quantitatively the border between truly nonlinear and quasi-linear dynamic behavior.

A particular property of the Morris-Lecar model is the delay in damping occurrence (similar to the so-called \textit{delayed loss of stability} \cite{Izh07}), which is especially pronounced when the stimulating current $I_{stim}$ is just slightly above $I_{max}$. During the delay, which can last, for example, 30 oscillation periods, it is practically impossible to distinguish the system dynamics from the case of periodic spike generation occurring at $I_{min}<I_{stim}<I_{max}$. In turn, the ordinary bifurcation analysis, which implies exponential dynamics, does not capture such a delay. The obtained formulas can be helpful for studying the limits of applicability of the bifurcation analysis in the considered case.

\bigskip

\textbf{Acknowledgments}

\bigskip

This work was partially funded by the Russian Foundation for Basic Research according to the research project \# 17-29-07093.

\newpage

\pagenumbering{gobble} 

\begin{figure*}[!th]
\centering
\includegraphics[width=1.0\textwidth]{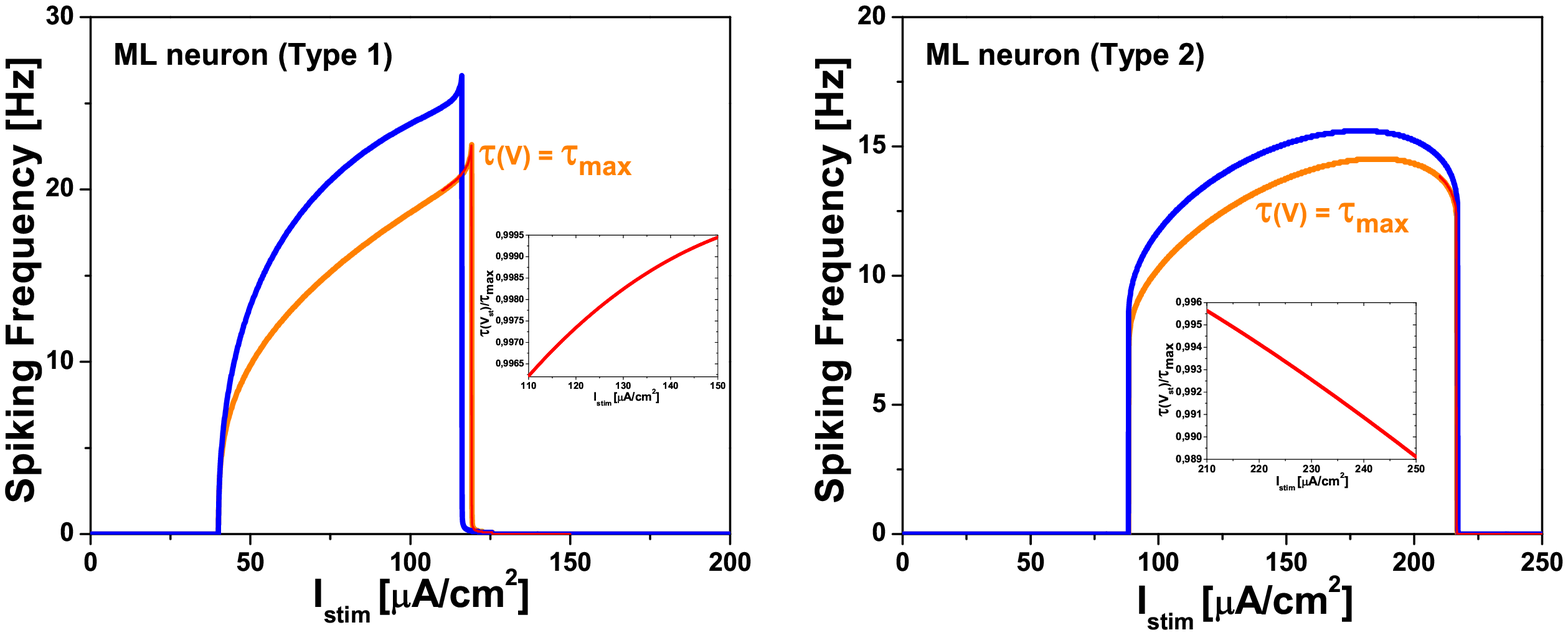}
\end{figure*}

\noindent \textbf{Figure S1}. Dependence of spike generation frequency (determined as the number of spikes divided by the time interval of 20000 ms) on constant stimulating current $I_{stim}$ for the Morris-Lecar (ML) model with the 1st (left graph) and 2nd (right graph) excitability types. The blue curves correspond to the standard ML model described in Sec. 2 of the main text. The orange curves correspond to a simplified version of the ML model with function $\tau(V)$, see Eq. (5), taken as a constant equal to its maximal value $\tau_{\max}$. In turn, the red curves near the upper boundary of the sustained spiking interval, which are virtually superimposed on the orange ones, correspond to the similar case where function $\tau(V)$ is also taken as a constant equal to $\tau(V_{st})$. The value $V_{st}$ is determined from equation $I_{ion}(V_{st},w_{\infty}(V_{st}))=I_{stim}$ and, above the upper boundary, $V_{st}$ corresponds to the stationary asymptotic value of the neuron potential. Finally, the inset in each graph shows the dependence of $\tau(V_{st})$ on $I_{stim}$ that is practically negligible (nevertheless, note that it is opposite for type 1 and type 2) so that one can safely use universal approximation $\tau(V)=\tau_{\max}$.

\noindent The parameters of the ML model with the excitability type 1 were as follows: $C_{m}=20$ $\mu$F/cm$^{2}$, $g_{Ca}=4$ mS/cm$^{2}$, $g_{K}=8$ mS/cm$^{2}$, $g_{L}=2$ mS/cm$^{2}$, $V_{Ca}=120$ mV, $V_{K}=-84$ mV, $V_{L}=-60$ mV, $V_{1}=-1.2$ mV, $V_{2}=18$ mV, $V_{3}=12$ mV, $V_{4}=17.4$ mV, $\tau_{\max}=14.925$ ms. These parameters result in the following values for the lower and upper boundaries of the sustained spiking interval of $I_{stim}$: $I_{min}=40$ $\mu$A/cm$^{2}$ and $I_{max}=116.1$ $\mu$A/cm$^{2}$.

\noindent In turn, the ML model of the excitability type 2 had the following parameters: $g_{Ca}=4.4$ mS/cm$^{2}$, $V_{3}=2$ mV, $V_{4}=30$ mV, $\tau_{\max}=25$ ms, with all the rest parameters being the same as those for the type 1. The corresponding values for the lower and upper boundaries of the sustained spiking interval are $I_{min}=88.3$ $\mu$A/cm$^{2}$ and $I_{max}=216.9$ $\mu$A/cm$^{2}$.

\newpage

\pagenumbering{gobble} 

\begin{figure*}[!th]
\centering
\includegraphics[width=1.0\textwidth]{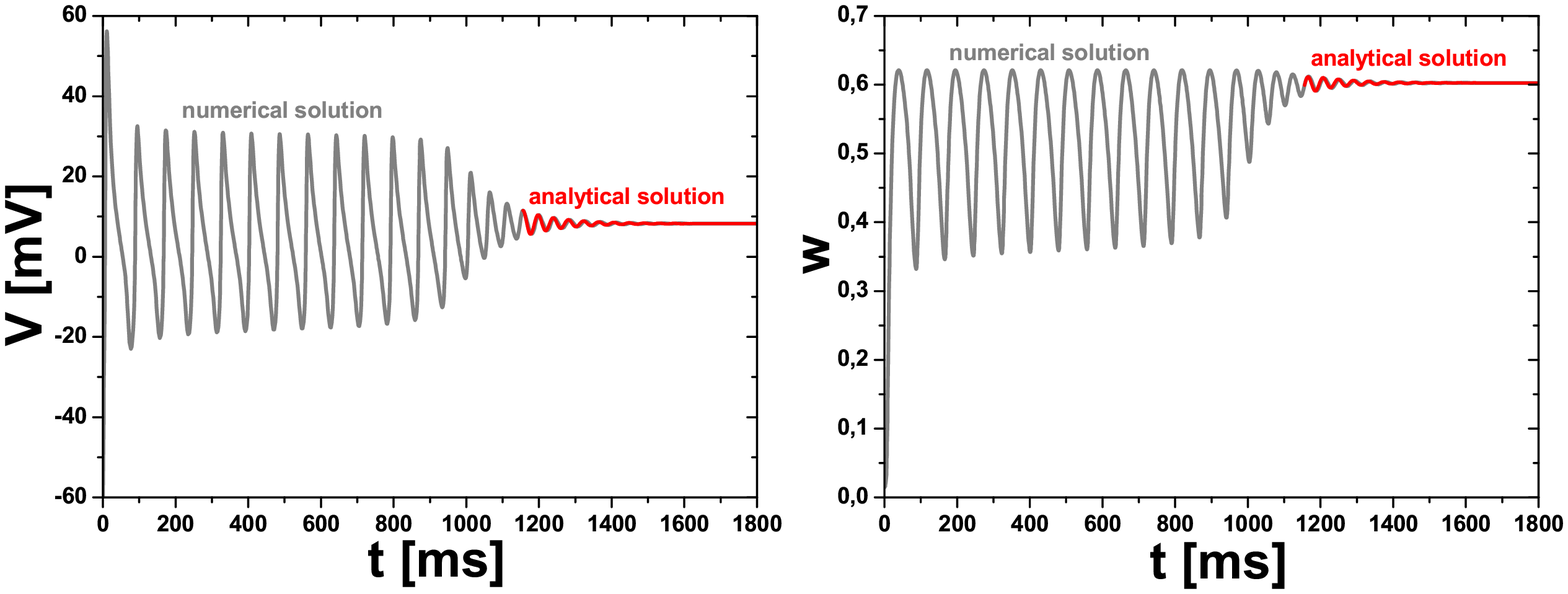}
\end{figure*}

\noindent \textbf{Figure S2}. Left graph: The gray curve is a numerical solution for dynamics of neuron potential $V(t)$ in the Morris-Lecar (ML) model with the 2nd excitability type at $I_{stim}$ = 216.995 $\mu$A/cm$^{2}$ $>$ $I_{max}$ = 216.9 $\mu$A/cm$^{2}$, where $I_{max}$ is the upper boundary of the sustained spiking interval (see the right graph in Fig. S1). The red curve is the analytical solution of the linearized system of equations of the ML model with initial conditions taken at the point of a local maximum of the potential ($t_{0}$ = 1156 ms, $V_{0}$ = 11.49 mV). Right graph: The corresponding numerical (gray) and analytical (red) solutions for $w(t)$. Parameters of the analytical formulas for this example are as follows: $V_{st}$ = 8.25 mV, $a = 0.6$, $\omega_{0}$ = 151.2 Hz, $\gamma$ = 9.76 Hz, $\omega$ = 150.9 Hz, $1/\tau$ = 40.2 Hz, $\eta = 0.065$, $\chi = 0.2$, $2\gamma/|A_{K}| = 14.43$, $U_{0}$ = 3.24 mV, $a/w_{0} = 1.0$, $W_{a} = 0.014$, and $W_{c} = -0.005$, where the last five parameters depend on $t_{0}$ and $V_{0}$ values.

\noindent The ML model parameters for the 2nd neuronal excitability type were as follows: $C_{m}=20$ $\mu$F/cm$^{2}$, $g_{Ca}=4.4$ mS/cm$^{2}$, $g_{K}=8$ mS/cm$^{2}$, $g_{L}=2$ mS/cm$^{2}$, $V_{Ca}=120$ mV, $V_{K}=-84$ mV, $V_{L}=-60$ mV, $V_{1}=-1.2$ mV, $V_{2}=18$ mV, $V_{3}=2$ mV, $V_{4}=30$ mV, $\tau_{\max}=25$ ms.

\end{document}